\documentclass[cits]{PoS}
\usepackage[utf8]{inputenc}

\title{Evaluation of disconnected contributions using GPUs}
\ShortTitle{Evaluation of disconnected contributions using GPUs}

\date{\today}
\author{C. Alexandrou$^{ab}$, V. Drach$^c$, K. Hadjiyiannakou$^b$, K. Jansen$^c$, G. Koutsou$^a$, A. Strelchenko$^a$ and \speaker{A.~Vaquero}$^a$ \\
\llap{$^a$} Computation-based Science and Technology Research Center (CaSToRC), The Cyprus Institute, \\
20 Constantinou Kavafi Street Nicosia 2121, Cyprus \\
\llap{$^b$} Departament of Physics, University of Cyprus, P.O. Box 20537, 1678 Nicosia, Cyprus \\
\llap{$^c$} NIC, DESY, Platanenallee 6, D-15738 Zeuthen, Germany}

\abstract{We calculate  on GPUs the disconnected diagrams associated with the nucleon form factors and moments of generalized parton distributions
using Nf=2+1+1 twisted mass fermions. We employ the truncated solver method (TSM) for estimating the
all-to-all propagators. Due to the fact that the TSM involves many low precision stochastic estimators, the usage of GPUs is essential
to perform efficiently the contractions and the inversions.}

\FullConference{The 30$^th$ International Symposium on Lattice Field Theory \\
Cairns Convention Centre, Cairns, Australia \\
Sunday, June 24 — Friday, June 29 2012}

\begin{document}

\section{Introduction}


The evaluation of disconnected diagrams is of paramount importance for eliminating systematic errors in the determination of proton and neutron observables. These contribute significantly in the evaluation of the $\eta'$
mass and strange content of the nucleon, and require a non-perturbative evaluation involving all-to-all propagators at a given time slice. This, and the inherent gauge noise associated with fermionic loops, explains why
most hadron studies neglect these contributions.


Fortunately, in the recent years there has been progress in algorithms and an increase in computational power, making these computations feasible.
On the algorithmic side, the introduction of improvements as the one-end trick, and the truncated solver method (TSM) had led to a significant reduction in the variance of disconnected computations. Using the properties 
of twisted mass fermions, one can further reduce the variance in isoscalar quantities by taking appropriate combinations of two flavors of twisted mass fermions.
On the hardware side, GPU units provide a large speed-up in the evaluation of quark propagators and contractions. For the TSM, they provide an optimal platform for swiftly increasing the amount of measurement 
we can perform.

\section{Methods for disconnected calculations}
\subsection{Stochastic estimation\label{stoSec}}

The exact computation of all-to-all propagators for the lattice volumes of physical interest is outside the current computer power. The fermionic matrix size ranges from $\sim10^6$ to
$\sim10^9$ in the largest volumes, thus an exact computation of the inverse would require an equal number of inversions, and the situation for timeslice-to-all propagators is equally unfeasible. A way to make
progress is to compute an unbiased stochastic estimation of the propagator \cite{Stochastic}: we generate a set of $N$ sources $\left|\eta_j\right\rangle$ randomly, by filling each component of the source with a number,
in our case a particular representation of the $\mathbb{Z}_2$ or $\mathbb{Z}_4$ group. Then the sources have the following properties:

\begin{equation}
\frac{1}{N}\sum_{j=1}^N\left|\eta_j\right\rangle = O\left(\frac{1}{\sqrt{N}}\right),\qquad\frac{1}{N}\sum^N_{j=1}\left|\eta_j\right\rangle\left\langle\eta_j\right| = \mathbb{I} + 
O\left(\frac{1}{\sqrt{N}}\right).
\label{eq1}
\end{equation}
The first property ensures that our estimation is unbiased. The second one allows us to reconstruct the inverse matrix by solving for $\left|s_j\right\rangle$ in 
$M\left|s_j\right\rangle = \left|\eta_j\right\rangle$ and calculating
\begin{equation}
M_E^{-1}:=\frac{1}{N}\sum_{j=1}^N\left|s_j\right\rangle\left\langle\eta_j\right|\approx M^{-1}.
\label{estiM}
\end{equation}
\noindent This way the computation becomes feasible, although it is still expensive due to the high number of inversions required to achieve a good estimate of $M^{-1}$ in Eq.~(\ref{estiM}).

The deviation of the estimator from the exact solution is given by
\begin{equation}
M^{-1}-M_E^{-1} = M^{-1}\times\left(\mathbb{I}-\frac{1}{N}\sum_{j=1}^N\left|\eta\right\rangle\left\langle\eta\right|\right).
\label{errS}
\end{equation}
From Eq. ~(\ref{eq1}) it is clear that the more stochastic sources are used, the smaller the stochastic error becomes. In fact, from Eq.(\ref{eq1}) and (\ref{errS}) we learn that the
errors decrease as $O\left(\frac{1}{\sqrt{N}}\right)$, as expected. Since we also have to deal with the gauge error, we would like to
minimize the statistical error by increasing the number of stochastic sources $N$ until we reach the gauge noise. For some cases, this may result in a large value of $N$, an expensive choice.

\subsection{The Truncated Solver Method}

The Truncated Solver Method (TSM) \cite{TSM} increases $N$ at a reduced cost by aiming at a low precision (LP) estimation of the inverse $\left|s_j\right\rangle_{LP} = \left(M^{-1}\right)_{LP}\left|\eta_j\right\rangle$,
where the inverter is truncated at reduced accuracy. The truncation criterium can be a large residual or equivalently a fixed number of iterations. This way we can increase the number of sources $N_{LP}$ cheaply, but we are introducing a bias
in our estimate due to the truncation. We correct the bias stochastically, by inverting a number of sources to high and low precision and taking the difference:

\begin{equation}
M_{E_{TSM}}:=\underbrace{\frac{1}{N_{HP}}\sum_{j=1}^{N_{HP}}\left[\left|s_j\right\rangle_{HP} - \left|s_j\right\rangle_{LP}\right]\left\langle\eta_j\right|}_{Correction} +
\underbrace{\frac{1}{N_{LP}}\sum_{j=N_{HP}}^{N_{HP}+N_{LP}}\left|s_j\right\rangle_{LP}\left\langle\eta_j\right|}_{Biased\quad estimate},
\label{estiTSM}
\end{equation}
which requires $N_{HP}$ high precision inversions and $N_{HP}+N_{LP}$ low precision inversions. If the convergence of the solver is fast, we only need a few high precision inversions to estimate properly the correction, and then
the error falls as $O\left(\sqrt{1/N_{LP}}\right)$. Therefore we want to ensure a good convergence for the solver; in our case this is ensured by the twisted mass regularization, which introduces a lower bound for the eigenvalues of the 
dirac operator.

The TSM needs tuning of its parameters, namely the precision of the LP  inversions and $N_{HP}/N_{LP}$ ratio, to get a safe result with maximum performance. For the first parameter we chose values already
used in the literature, i.e., the residual $\rho_{LP}\sim 10^{-2}$ \cite{Alexandrou:2012zz}. The tuning of the second parameter was performed empirically: we took a disconnected diagram we expected to yield a large stochastic error, and we
optimized $N_{HP}$ and $N_{LP}$ so as to get the minimum error at the lowest computer cost. As shown in Fig. \ref{PerfPlots}, the error decreases as the number of HP or LP increases. A good compromise for this particular diagram is $HP=12$ and $LP=300$ as the cheapest point that saturates to the gauge noise. Since the tuning depends on the diagram to be computed, we decided to take the more conservative number of 24 for the number of HP sources.

\subsection{The one-end trick}

The properties of the twisted mass action provide a powerful method to reduce the variance of the disconnected diagrams. The standard way to compute the disconnected diagrams is to generate $N$ stochastic sources $\eta_r$, invert them, and compute the diagrams
corresponding to operator $X$ as $\frac{1}{N}\sum^N_r \left\langle \eta^\dagger_r X s_r\right\rangle \approx \textrm{Tr}\left(M^{-1}X\right)$, where the operator $X$ is expressed in the twisted basis. However, if the operator $X$
involves an isovector combination in the twisted basis, one can resort to the identity $M_u - M_d = 2i\mu a\gamma_5$, which becomes $M^{-1}_u - M^{-1}_d = -2i\mu aM_d^{-1}\gamma_5M_u^{-1}$ for the propagators:

\begin{equation}
\frac{2i\mu a}{N}\sum^N_r \left\langle s_r \gamma_5 X s_r\right\rangle = \textrm{Tr}\left(M_u^{-1}X\right) - \textrm{Tr}\left(M_d^{-1}X\right) + O\left(\frac{1}{\sqrt{N}}\right).
\label{loopVv}
\end{equation}
As a result of this substitution, the fluctuations are reduced by the small $\mu$ factor. Most important is the implicit sum of $V$ terms in the product $M_d^{-1}\gamma_5 M_u^{-1}$.
The difference of propagators exhibits a signal-to-noise ratio of $1/\sqrt{V}$, but in the product it becomes $V/\sqrt{V^2}$. In fact, a comparison between the two methods reveals a
large reduction in the errors at the same computer cost \cite{vv1, vv2, Dinter:2012tt}. The drawback of this technique is its inapplicability to operators lacking a $\tau_3$ flavour matrix in the twisted basis. A generalized
version of the trick can be developed from the identity $M_u + M_d = 2D_W$, with $D_W$ the Dirac-Wilson operator without a twisted mass term. After some algebra,

\begin{equation}
\frac{2}{N}\sum^N_r \left\langle s_r \gamma_5 X\gamma_5 D_W s_r\right\rangle = \textrm{Tr}\left(M_u^{-1}X\right) + \textrm{Tr}\left(M_d^{-1}X\right) + O\left(\frac{1}{\sqrt{N}}\right),
\label{loopStD}
\end{equation}
but the lack of the $\mu$ suppresion factor introduces a considerable penalty in the signal-to-noise ratio.

\section{Simulation details}

In order to test these methods, we analyzed 4698 configurations of the $B55$ ensemble of the ETMC collaboration. This ensemble is a $32^3\times 64$ lattice and was generated with $2+1+1$ dynamical fermions,
at pion mass  $m_\pi \approx 360$ MeV and strange and charm quark masses fixed at about their physical values. The resulting lattice spacing is $a = 0.086(1)$ fm
determined from the nucleon mass resulting in $m_{\pi}L \sim 5$.
The disconnected diagrams were computed by making intensive use of a modified version of the QUDA library \cite{QUDA1,QUDA2}, which implemented new code and kernels to do the required
inversions and contractions on the GPUs. For the Fourier transform we used the CUFFT library.

\begin{figure}[h!]
\begin{center}
$\begin{array}{cc}
\includegraphics[width=0.42\textwidth,angle=0]{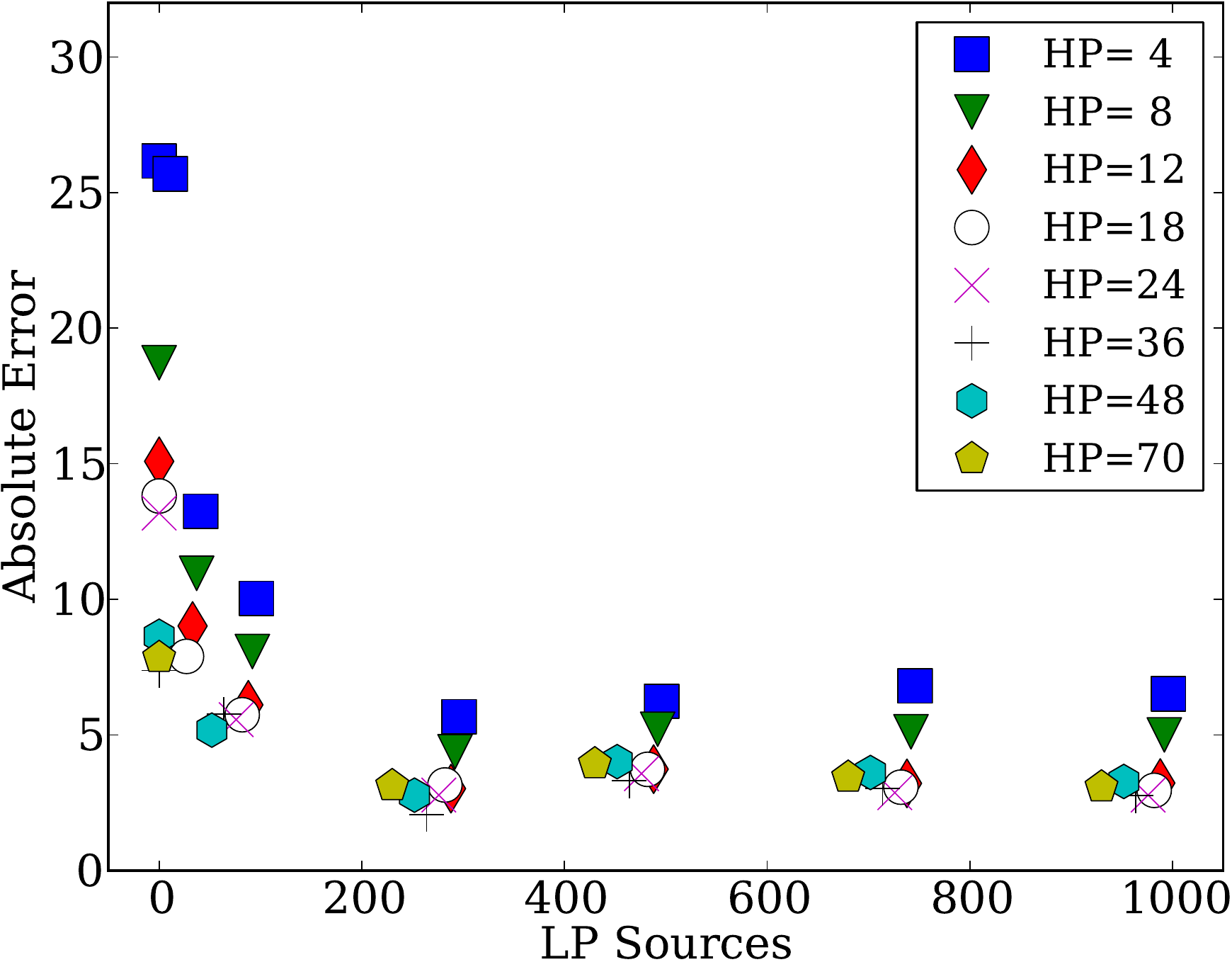} &
\includegraphics[width=0.42\textwidth,angle=0]{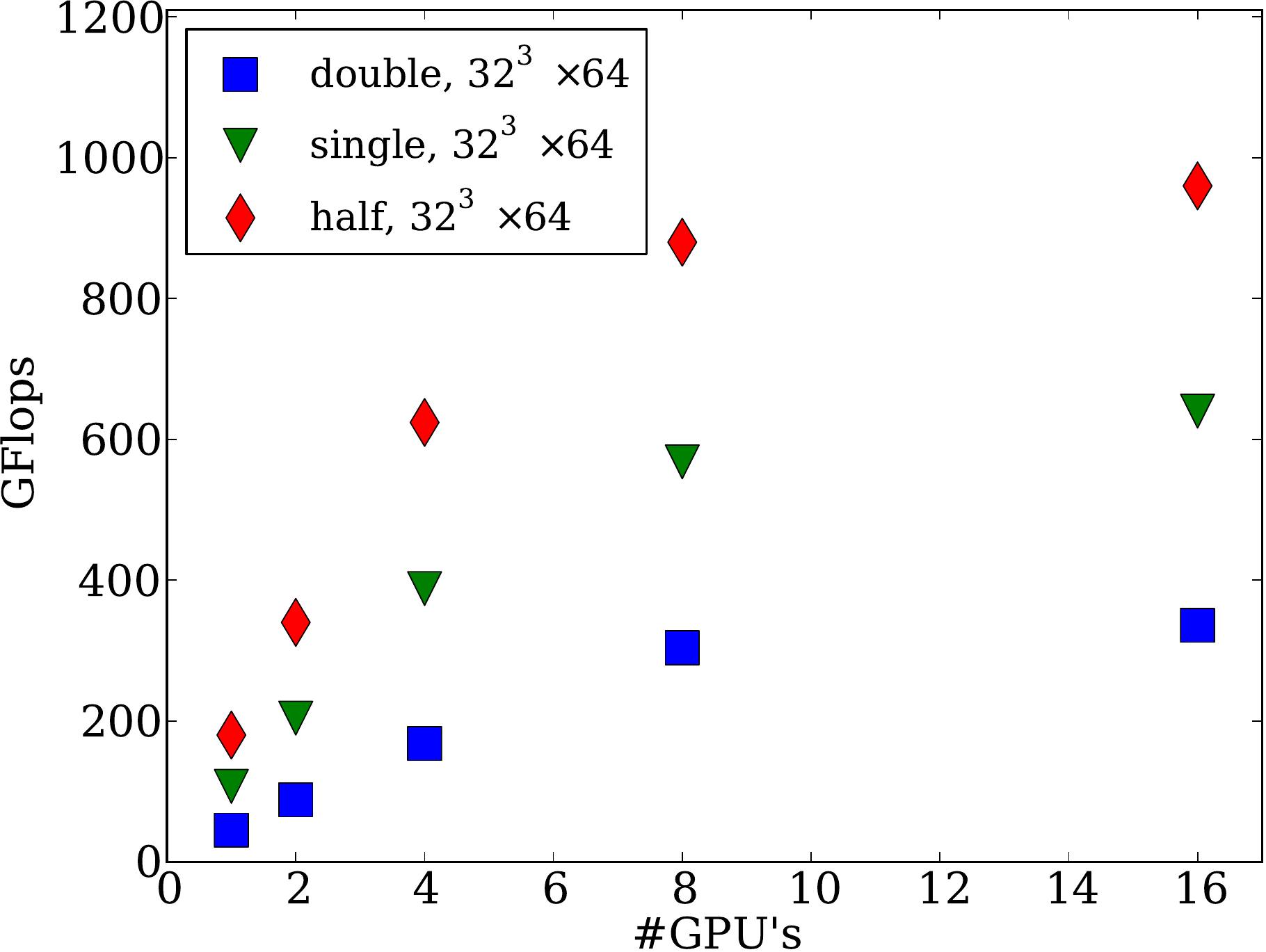}
\end{array}$
\caption{Left: Tuning of the number of HP and LP stochastic noise vectors for the TSM using 50 configurations of the B55.32 ensemble for the the traceless version of the
operator $i\bar\psi\gamma_3 D_3\psi$ at a given
value of the insertion time $t_i = 8$ and sink time $t_s = 16$. The error is
shown versus $N_{LP}$ for different values of $N_{HP}$ marked by the different ploting symbols given in the legend. Right: Strong scaling of the multi-GPU code
for this ensemble.\label{PerfPlots}}
\end{center}
\end{figure}

The QUDA library allowed for multi-GPU calculations, so 2 GPUs worked in parallel per configuration. As seen in the right graph of Fig. \ref{PerfPlots}, the scaling for a few GPUs is very good, with a $\sim90\%$
increase in performace when adding the second GPU. This result holds up to 8 GPUs, where there is a drop, beyond that the advantages of adding new GPUs are only useful in the case of lack of memory. It is
remarkable however that we can reach TFlop sustained performance with just a few GPUs.

The computations were performed on GPU clusters with NVidia fermi GPUs, mainly Tesla M2070 with 6Gb of memory, but also Tesla M2090 and M2050. The noise sources were generated on-the-fly, and the propagators were not stored, in order to save storage and I/O time.

\section{The analysis with the summation method}

One of the advantages of the one-end trick for twisted mass fermions is the fact that, since the noise sources must be on all sites, we obtained results for all the possible insertions for free. This feature
enables us to use the summation method to perform the ratio analysis.

The method is known since a long time \cite{Sum1, Sum2, Sum3}, and requires the knowledge of the three point function for all possible insertion times. The advantage is the reduction of the noise due to the excited states by an exponential of 
the sink time, $e^{-kt_s}$, as opposed to the standard decrease with the insertion time $e^{-Kt_i}$. In this method we sum, for every value of $t_s$, the ratios from $t_i = 0$ up to $t_i = t_s$, $R_{Sum}(t_s) = 
\sum_{t_i=0}^{t_i=t_s} R_{Plateau}(t_i,t_s)$. Thence the dependence of the ratio on $t_i$ dissappears. The ratio $R_{Plateau}$, computed as the quotient between the three-point function and the two-point function,
can be written as $R_{Plateau}(t_i,t_s) = R_{GS} + O(e^{-Kt_i}) + O(e^{-K't_s})$, where $R_{GS}$ is the uncontaminated ratio, and the other contributions are the undesired excited states. After performing the sum in
$t_i$, we get  the ratio as a slope $R_{Sum}(t_s) = t_sR_{GS} + c(K,K') + O(e^{-Kt_s}) + O(e^{-K't_s})$, and the contributions of the excited states become a geometrical series in $t_i$ whose sum
decays as $t_s$. Therefore we expect a better suppression of the excited states at the same $t_s$. The drawback is that we now need to fit to a straight line with two fitting parameters instead of one.

\section{Results}

\begin{figure}[h!]
\begin{center}$
\begin{array}{cc}
\includegraphics[width=0.42\textwidth,angle=0]{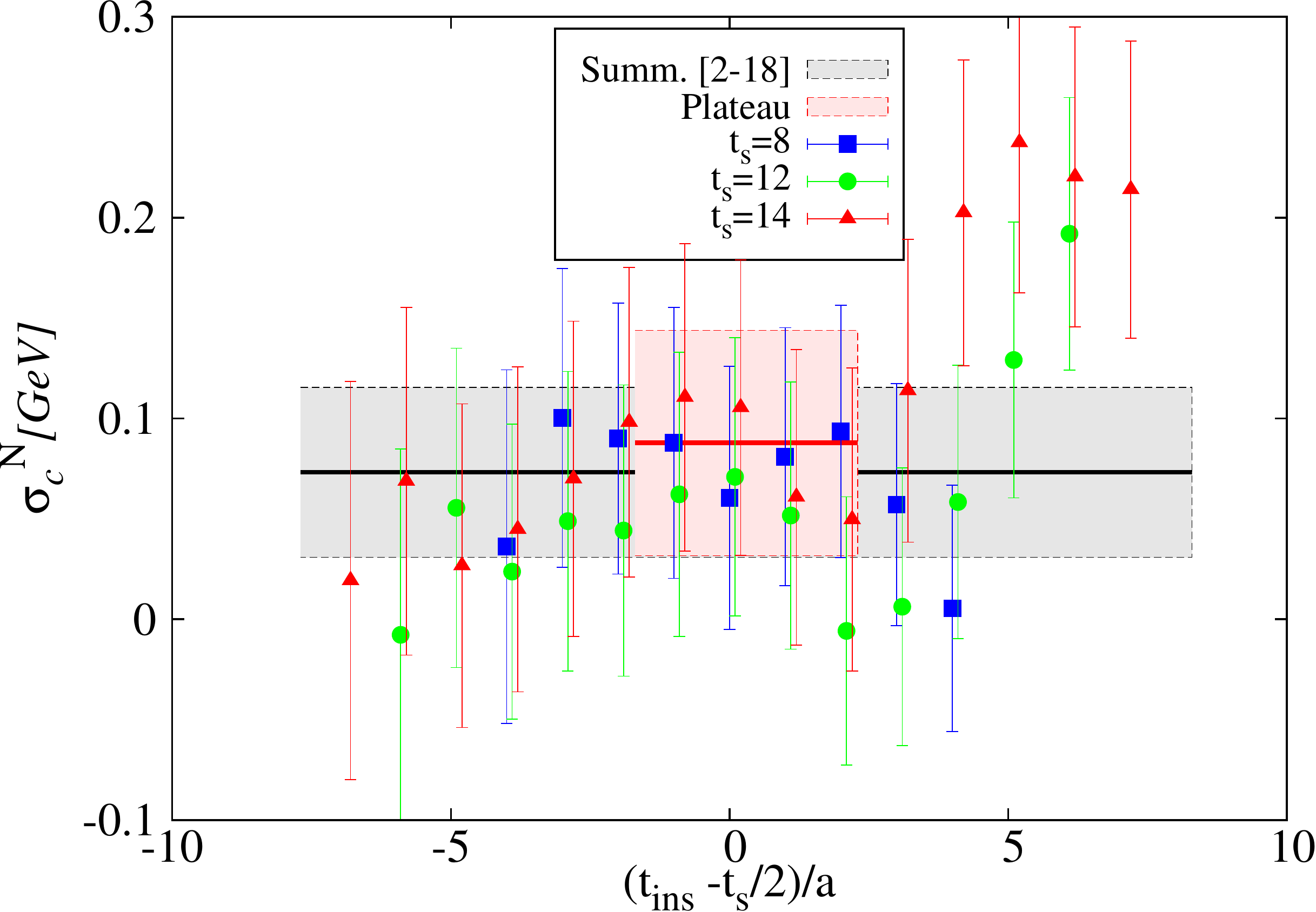} &
\includegraphics[width=0.42\textwidth,angle=0]{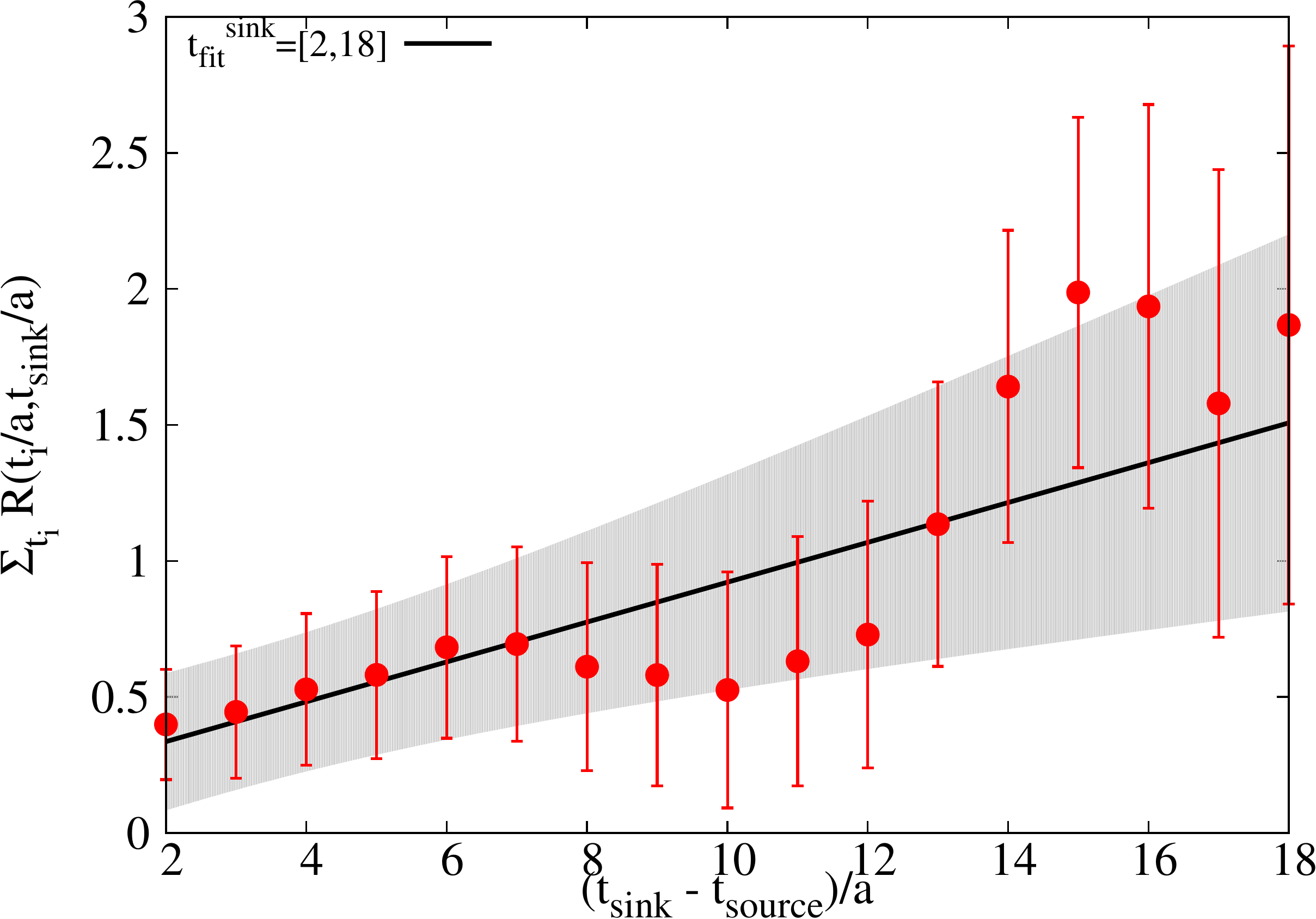}
\end{array}$
\caption{Left: Charm content for the nucleon, from $R_{Plateau}(t_i,t_s)$. The grey band is the value obtained from the summation method (right).\label{CharmNucleon}}
\end{center}
\end{figure}

\begin{figure}[h!]
\begin{center}$
\begin{array}{cc}
\includegraphics[width=0.42\textwidth,angle=0]{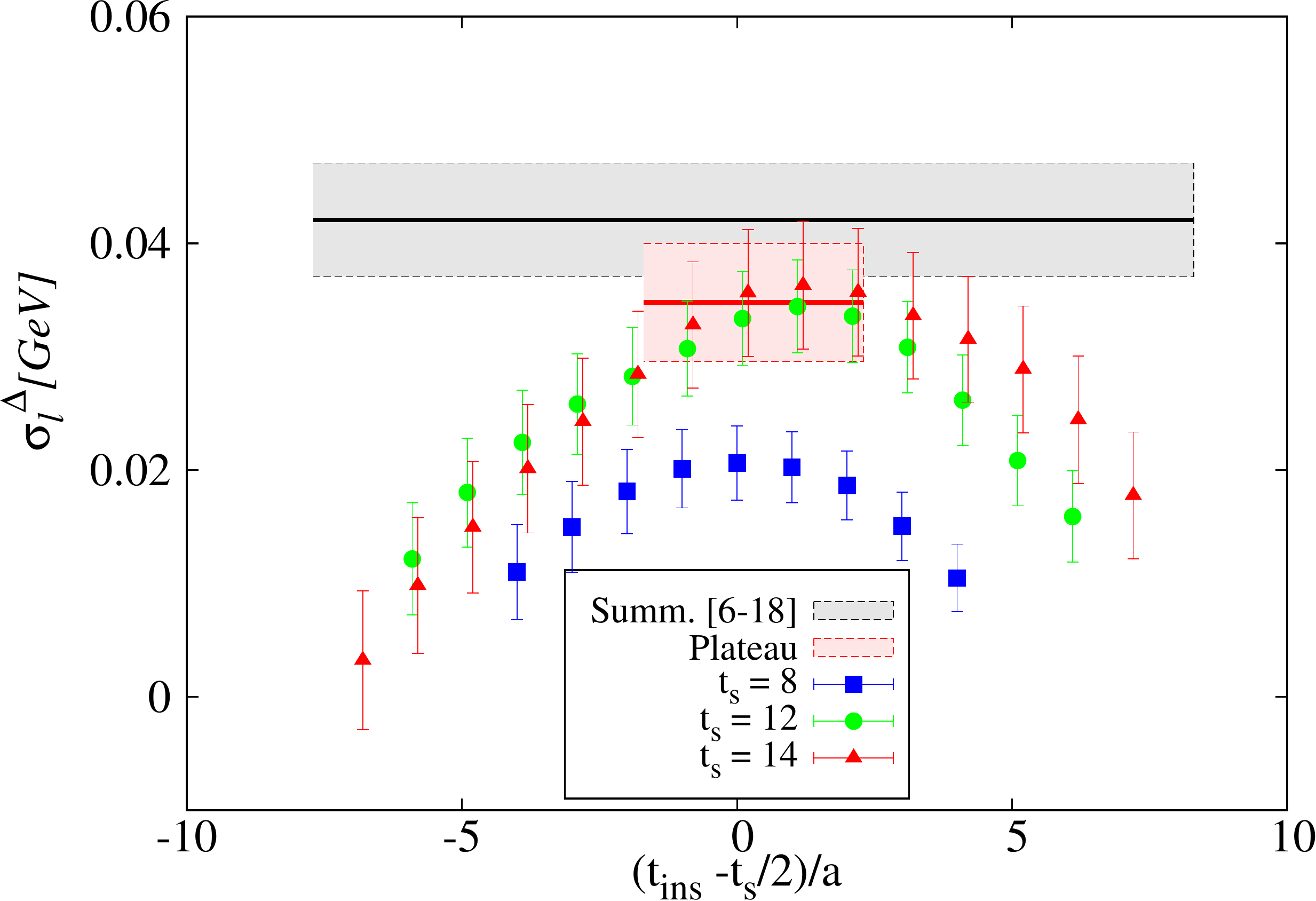} &
\includegraphics[width=0.42\textwidth,angle=0]{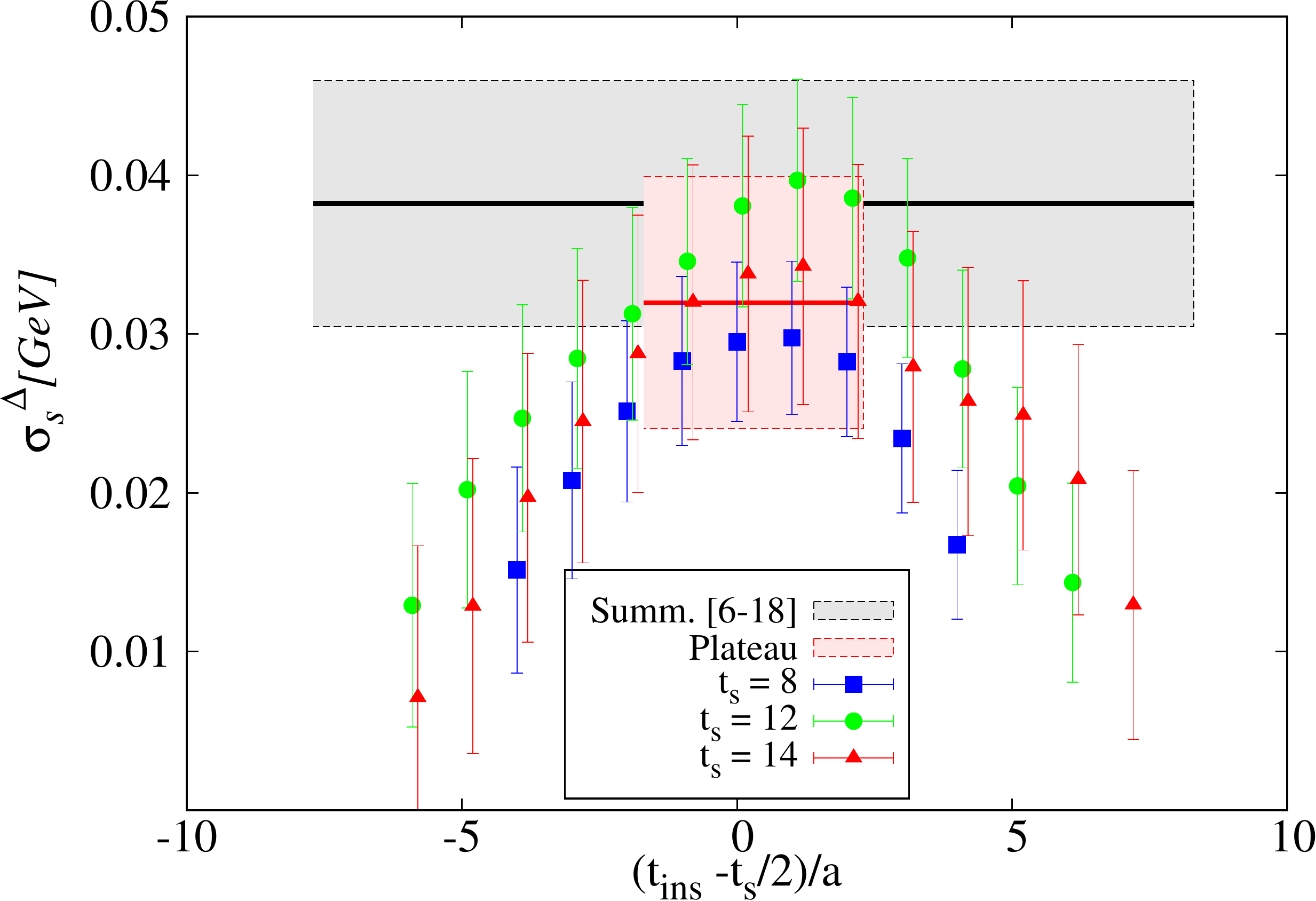}
\end{array}$
\caption{Left: Disconnected contribution to the light $\sigma$-term of the $\Delta$ from $R_{Plateau}(t_i,t_f)$. Right: The strange $\sigma$-term of the $\Delta$.\label{Delta}}
\end{center}
\end{figure}

\begin{figure}[h!]
\begin{center}$
\begin{array}{cc}
\includegraphics[width=0.42\textwidth,angle=0]{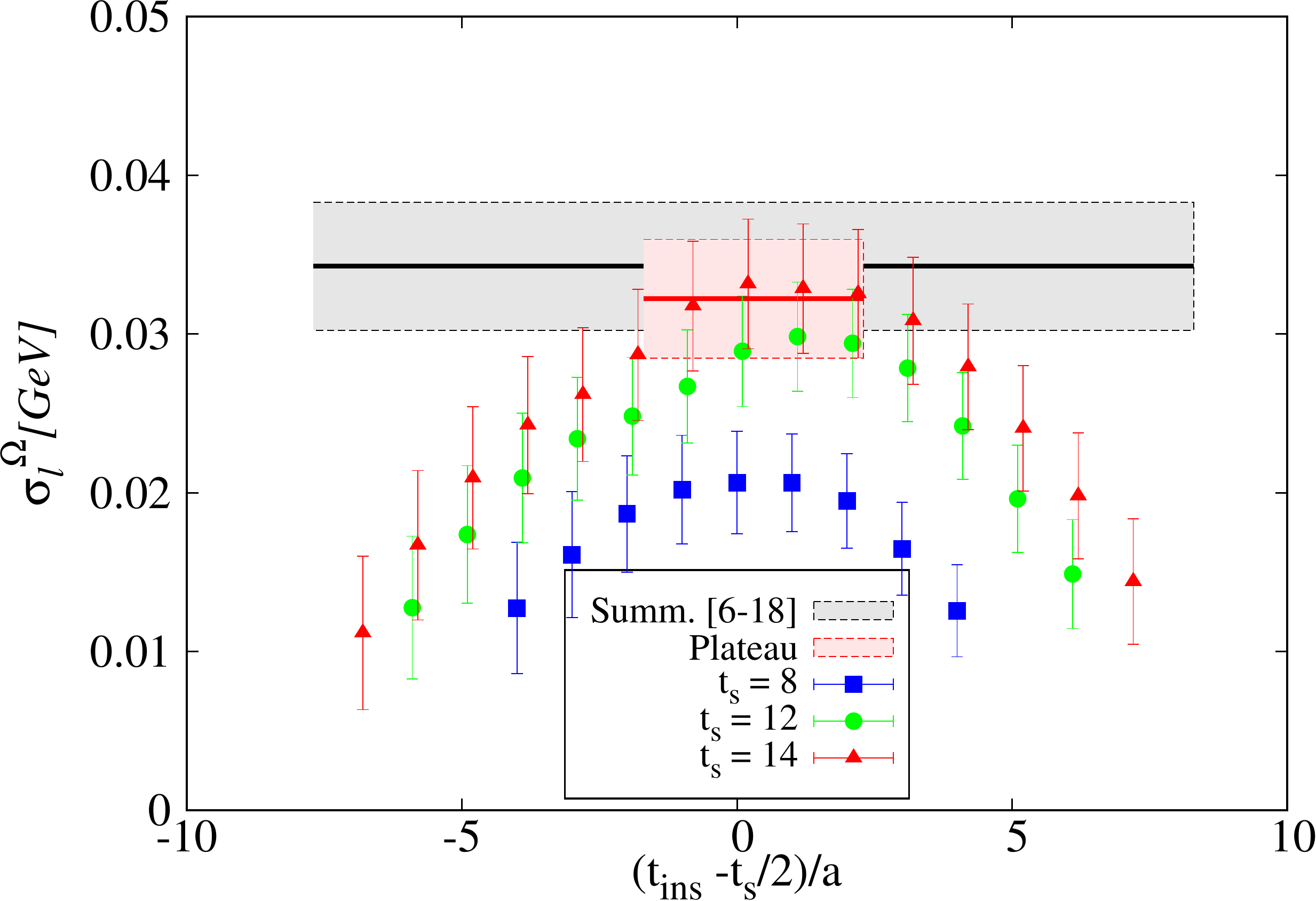} &
\includegraphics[width=0.42\textwidth,angle=0]{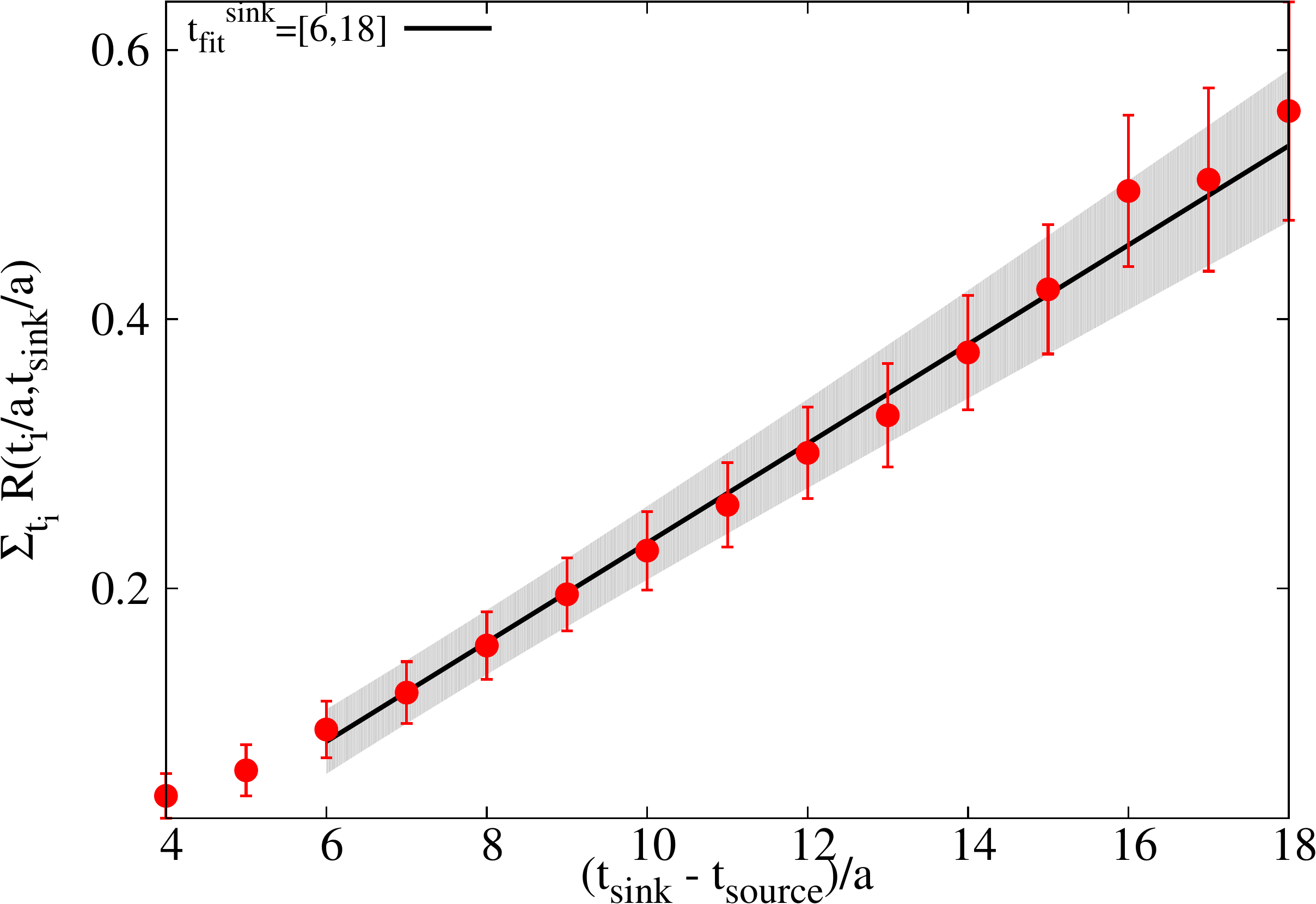}
\end{array}$
\caption{Left: Light $\sigma$-term of the $\Omega$ from $R_{Plateau}(t_i,t_s)$. The grey band is the value obtained from the summation method by fitting the slope shown on the right.\label{LightOmega}}
\end{center}
\end{figure}

\begin{figure}[h!]
\begin{center}$
\begin{array}{cc}
\includegraphics[width=0.42\textwidth,angle=0]{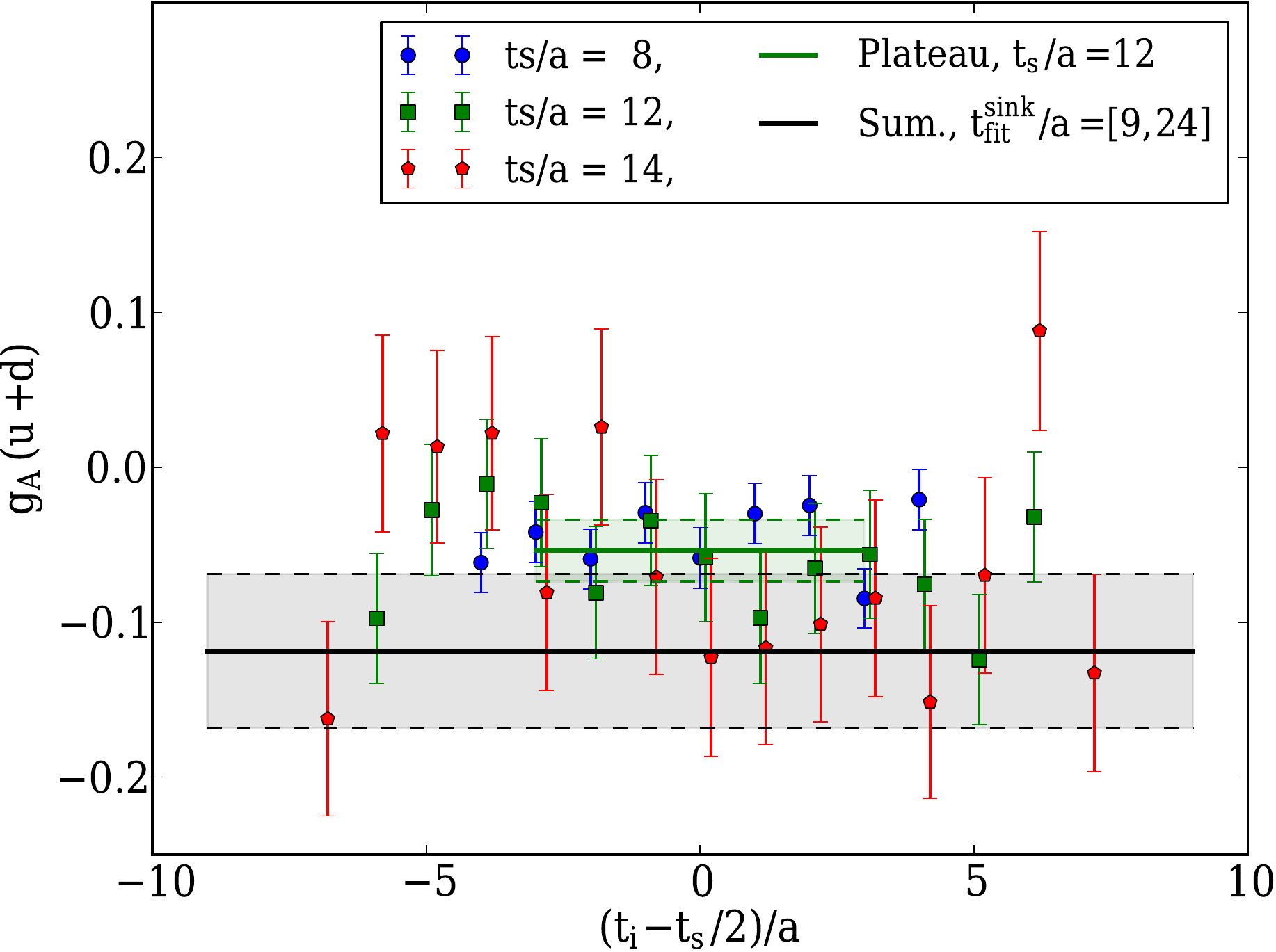} &
\includegraphics[width=0.42\textwidth,angle=0]{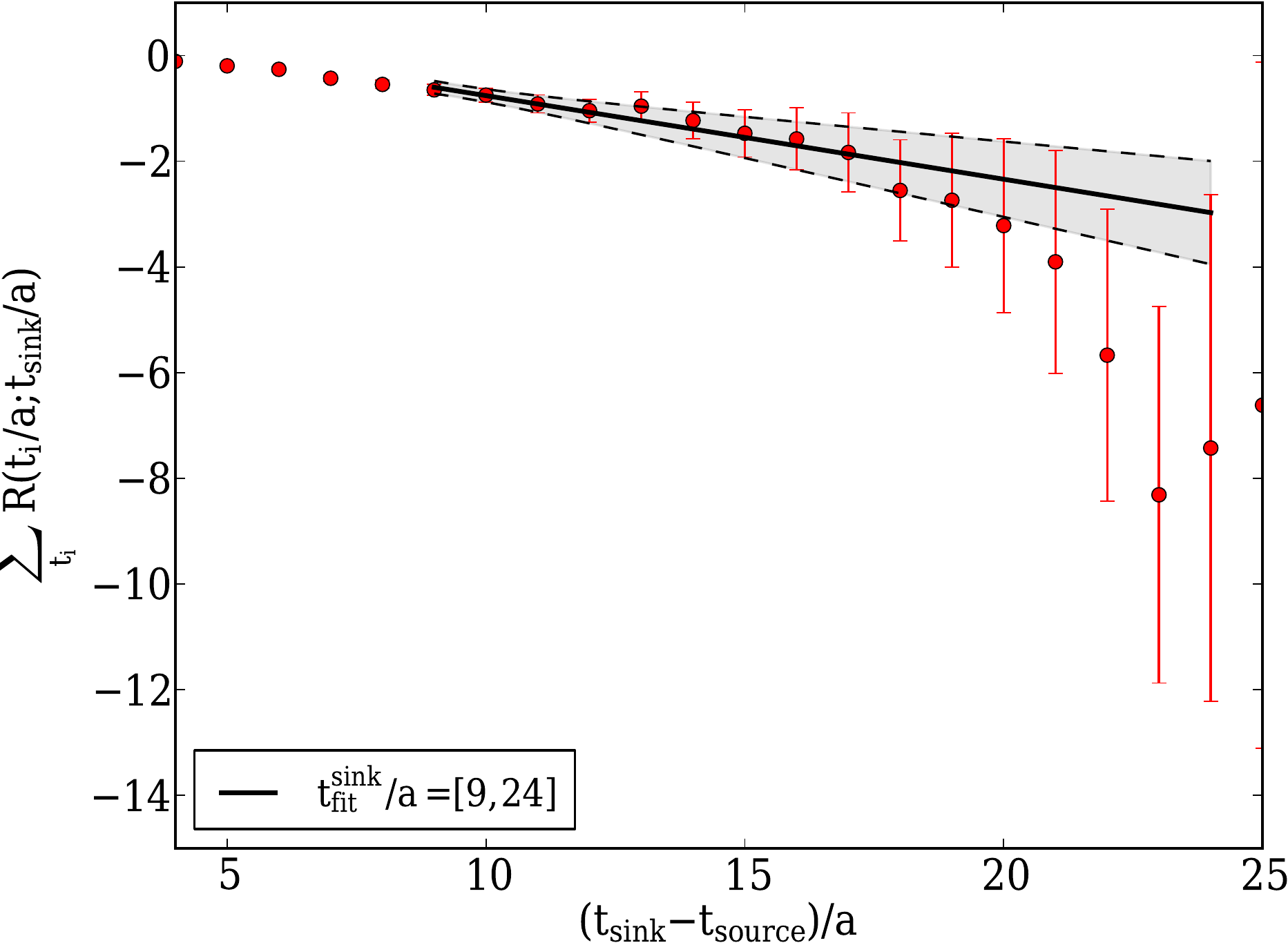}
\end{array}$
\caption{Disconnected contribution to the isoscalar $g_A$ using the generalized one-end trick of Eq.({\protect \ref{loopStD}}). The results are noisier than those obtained for operators calculated using 
the standard one-end trick of Eq.({\protect \ref{loopVv}}).\label{gALight}}
\end{center}
\end{figure}

We combined the GPU-computed diagrams with nucleon 2-point functions in order to get the ratios for $g_A$ and $\sigma_c^N$. Each disconnected diagram was combined with a set of 5 2-point functions, with randomized
positions for each one of the 2912 configurations, where the 2-point functions were computed for proton and neutron, propagating backwards and forwards. In this manner we produced $20$ measurements 
per gauge configuration. The slope obtained in the summation method changes as the sink-source separation increases and fitting too early would yield a wrong result. The two methods give consistent results,
and therefore combining both one can ensure that we have a large enough sink-source separation for excited states to be neglected.

For the $\sigma_c^N$ (Fig.\ref{CharmNucleon}), more statistics are needed to understand the change in slope in the summation method. In the summation of the $\Delta$ we observed a similar behaviour, but it was quite reduced and
the results agree with the plateaus (Fig.\ref{Delta}), even when our statistics were smaller: 4643 configurations combined with 4 different 2-point functions propagating forwards and backwards ($8$ measurements).
In contrast, the $\Omega$ (Fig.\ref{LightOmega}) yields a strong signal with the same statistics.

The generalized version of the one-end trick as expected is more noisy. Our results for the isoscalar nucleon axial charge, $g_A^{is}$ are shown in Fig.~\ref{gALight} and are in agreement with recent
evaluation using Clover fermions \cite{gAResult}.

\section{Conclusions}


The computation of disconnected contributions for flavour singlet quantities has become feasible, due to the improvement in the algorithms and to the increase in computational resources. In this work we show that we can get
reliable results for disconnected contributions to the $\sigma$-terms and the isoscalar axial charge. GPUs are particularly efficient for the evaluation of disconnected diagrams using the TSM, yielding a huge
improvement in the computation of LP inversions and contractions. In additon, the one-end trick allows a reduction of the variance at the same computational cost, as well as  getting the fermion loops for all the possible insertion
times for free. This property, together with the application of the plateau and the summation methods, as well as the generalized one-end trick, allowed us to compute nucleon observables wiere disconnected diagrams play an important role.


\section*{Acknowledgments}

A. Vaquero is supported by the Research Promotion Foundation (RPF) of Cyprus under grant $\Pi$PO$\Sigma$E$\Lambda$KY$\Sigma$H/NEO$\Sigma$/0609/16. Computations are performed on GPUs on Cy-Tera (Cyprus) supported by RPF under the grant NEA Y$\Pi$O$\Delta$OMH/$\Sigma$TPATH/0308/31,  Judge at Jülich Forschungzentrum (JSC) (Germany), Forge at NCSA Illinois (USA) and Minotauro at BSC (Spain) through PRACE. Forward propagators were computed on Jugene at JSC through PRACE Tier-0 access.

\end{document}